\documentclass[twocolumn,showpacs,amsmath,amssymb]{revtex4}
\usepackage{graphicx}
\usepackage{dcolumn}
\usepackage{bm}
\newcommand{\be}{\begin{eqnarray}}
\newcommand{\en}{\end{eqnarray}}
\newcommand{\ben}{\begin{eqnarray*}}
\newcommand{\enn}{\end{eqnarray*}}
\newcommand{\pa}{\partial}
\newcommand{\na}{\nabla}
\newcommand{\f}{\frac}
\newcommand{\td}{\tilde}
\newcommand{\m}[1]{\mathbb{#1}}

\newcommand{\p}{\paragraph{}}
\newcommand{\bi}{\begin{itemize}}
\newcommand{\ei}{\end{itemize}}

\newcommand{\la}{\langle}
\newcommand{\ra}{\rangle}

\newcommand{\R}{\Rightarrow}
\renewcommand{\O}{\Omega}
\renewcommand{\p}{\bot}

\begin{document}
\title{On the third order structure function for rotating 3D homogeneous turbulent flow}
\author{Sagar Chakraborty}
\email{sagar@bose.res.in}
\affiliation{S.N. Bose National Centre for Basic Sciences, Saltlake, Kolkata 700098, India}
\author{J.K. Bhattacharjee}
\email{tpjkb@iacs.res.in}
\affiliation{IACS, Theoretical Physics, 2A\&2B Raja S C Mullick Road, Kolkata 700032, India}
\date{April 25, 2007}
\begin{abstract}
A form for the two-point third order structure function has been calculated for three dimensional homogeneous incompressible slowly rotating turbulent fluid.
It has been argued that it may possibly hint at the initiation of the phenomenon of two-dimensionalisation of the 3D incompressible turbulence owing to rotation.
\end{abstract}
\pacs{47.27.–i, 47.27.Jv}
\maketitle
\section{INTRODUCTION}
\indent With possible realistic applications to flows in oceans and atmospheres, rotating turbulence attracts interests of oceanographers, geophysicists, mathematicians and physicists alike.
Rotation seems to be serving as a bridge between 2D, quasi-2D and 3D turbulences, a fact well-established with the discovery of two-dimensionalisation of 3D turbulence due to rotation.
\\
\indent In the steady non-turbulent flow, for low Rossby number ($Ro=U/2L\O$) and high Reynolds number ($Re=UL/\nu$), Taylor-Proudman theorem\cite{Batchelor} argues that rotation two-dimensionalises the flow.
This argument is often mistakenly extended to turbulent flows to explain the rotation induced two-dimensionalisation arising therein.
The two-dimensionalisation of the 3D turbulent flow in presence of rotation has begun to be understood as a subtle non-linear effect which is distinctly different from Taylor-Proudman effect.
\\
\indent Cambon {\it et al.}\cite{Cambon1} showed that in the presence of rotation, the transfer of energy from small to high wavenumbers is inhibited; at the same time, the strong angular dependence of this effect leads to a draining of the spectral energy from the parallel to the normal wave vectors (w.r.t. the rotation axis) showing a trend towards two-dimensionalisation.
\\
\indent Waleffe\cite{Waleffe} used helical decomposition of the velocity field to study the nature of triad interactions in homogeneous turbulence and coupling it with the instability assumption predicted a transfer of energy toward wave vectors perpendicular to the rotation axis under rapid rotation.
The helical decomposition turns out to be very handy to deal with rapidly rotating turbulent flow.
In that case the linear eigensolutions of the problem, the so-called inertial waves, have the structure of helical modes.
The assumption about the triadic transfers, coupled with resonance condition for non-linear interaction between inertial waves, show that there will be a tendency toward non-linear two-dimensionalisation of the flow.
\\
\indent Simulations by Smith {\it et al.}\cite{Smith} speak volumes for the two-dimensionalisation effect.
They showed the coexistence of inverse cascade (a typical feature of 2D turbulence) and forward cascade in forced rotating turbulence within a periodic box of small aspect ratio.
In the simulations, the ratio of the mean rates of energy dissipated to the energy injected decreased almost linearly, for $Ro$ less than a critical value, with decrease in $Ro$ (increase in angular velocity $|\vec{\O}|$).
By the way, a very recent numerical study\cite{Waite} shows similar transition from stratified to quasi-geostrophic turbulence, manifested by the emergence of an inverse cascade -- a conclusion that agrees with that of Lindborg\cite{Lindborg}.
\\
\indent Although recent experiments by Baroud {\it et al.}\cite{Baroud1,Baroud2} and Morize {\it et al.}\cite{Morize1,Morize2}  have shed some light on the two-dimensionalisation effect, the scaling of two-point statistics and energy spectrum in rotating turbulence remains a controversial topic.
Zhou\cite{Zhou} in analogy with MHD turbulence proposed an energy spectrum $E(k)\sim k^{-2}$ for rapidly rotating 3D turbulent fluid and this does seem to be validated by some experiments\cite{Baroud1,Baroud2} and numerical simulations\cite{Yeung,Hattori,Reshetnyak,Muller}.
But some experiments\cite{Morize1} do not tally with this proposed spectrum.
They predict steeper than $k^{-2}$ spectrum and this again seem to be drawing some support from numerical results\cite{Yang,Bellet} and analytical results found using wave turbulence theory\cite{Galtier,Cambon2}.
\\
\indent To be precise, if one wishes angular velocity to become a relevant parameter in constructing the energy spectrum $E(k)$, simple dimensional analysis would lead one to:
\be
E(k)\propto\O^{\frac{3m-5}{2}}\varepsilon^{\f{3-m}{2}}k^{-m}
\label{0}
\en
where $m$ is a real number.
$m$ should be restricted within the range 5/3 to 3 to keep the exponents of $\O$ and $\varepsilon$ in relation (\ref{0}) positive.
The two limits $m=5/3$ and $m=3$ corresponds to isotropic homogeneous 3D turbulence and 2D turbulence respectively.
The spectrum due to Zhou -- $E(k)\sim k^{-2}$ -- is due an intermediate value of $m=2$.
So, as far as the present state of the literature on rotating turbulence goes, two-dimensionalisation of 3D turbulence would mean the dominance of a spectrum which goes towards $E(k)\sim k^{-3}$ and which may choose to settle at $E(k)\sim k^{-2}$, an issue yet to be fully resolved.
\\
\indent These all studies are for low $Ro$ high $Re$ limit while high $Ro$ high $Re$ limit has been rather less ventured in relation to two-dimensionalisation of turbulence.
Now, the two-point third order correlation function ($S_3$) in homogeneous isotropic turbulence has a rare non-trivial exact result.
In this paper we have studied the two-point third order correlation function ($S_3$) in this latter regime of high $Ro$ high $Re$ and went on to argue that a spectrum which goes as $k^{-2.87}$ ({\it i.e.,} in between $k^{-2}$ and $k^{-3}$) exists although dominated by the 3D spectrum $E(k)\sim k^{-5/3}$ the reason for which of course is that we are dealing with low rotation limit.
This obviously hints at the initiation of the two-dimensionalisation effect with slow rotation.
\\
\indent Also, if one goes by the procedure given in the book by Frisch\cite{Frisch} to derive the form of the correlation function in $d$-D turbulence with the {\it assumption} of forward energy cascade, one would land up on\cite{Gawedzki}:
\be
S_3\equiv\left<\left[\left\{\vec{v}(\vec{x}+\vec{l})-\vec{v}(\vec{l})\right\}.\f{\vec{l}}{|\vec{l}|}\right]^3\right>=-\f{12}{d(d+2)}\varepsilon l
\label{1}
\en
where $\varepsilon$ is the mean rate of dissipation of energy per unit mass.
This result is not quite true for the two-dimensional case since it gives for $d=2$, $S_3=-(3/2)\varepsilon l$ and not $S_3=(3/2)\varepsilon l$ because the calculation doesn't take into account the conservation of enstrophy which causes the reverse cascade of energy\cite{Kraichnann}.
It might be noted that $S_3=(3/2)\varepsilon l$ for $d=2$ is for the regime of scales larger than the forcing scale\cite{Sagar,Bernard}.
If, using calculations of structure functions, in the limit of high $Ro$ and high $Re$, one wishes to see whether a trend towards two-dimensionalisation of 3D homogeneous isotropic turbulence occurs or not, then basically one would have to check (a) if $S_3=-(4/5)\varepsilon l$ at small scales for 3D turbulence shows a tilt towards $S_3=(3/2)\varepsilon l$ at large scales for the 2D turbulence and (b) if the forward energy cascade is depleted at the smaller scales.
As we shall show here, in the lowest order calculation this is what one may get, again hinting at the initiation of the effect of two-dimensionalisation of 3D turbulence owing to the small anisotropy induced by slow rotation.
\section{Relevant Scales in rotating turbulence}
Let us look in to the various length scales that have to be taken into consideration while talking about a homogeneous rotating turbulence which basically satisfies following version of Navier-Stoke's equation:
\be
\f{\pa\vec{v}}{\pa t}+\left(\vec{v}.\vec{\na}\right)\vec{v}&=&-\f{1}{\rho}\vec{\na}P-\vec{\Omega}\times\left(\vec{\Omega}\times\vec{x}\right)\nonumber\\
&&-2\vec{\Omega}\times\vec{v}+\nu\na^2\vec{v}+\vec{f}
\label{nse}
\en
Various parameters to be considered are: $\nu$ (kinematic viscosity), $\varepsilon$ (finite mean rate of dissipation of energy per unit mass), $\Omega$ (angular velocity) and $l_0$ (integral scale which typically is the system-size).
The three important time-scales involved in the system are: $t_l\sim\varepsilon^{-1/3}l^{2/3}$ (eddy-turnover time or circulation time for the eddy of scale $l$; $l\le l_0$), $t_{\Omega}\sim\Omega^{-1}$ and $t_d\sim l^2/\nu$ (diffusion time scale).
It is well-known that a length scale $l_\Omega=\sqrt{(\varepsilon/\Omega^3)}$ is what responsible for the estimation of the anisotropy introduced by the rotation.
The competition between the time-scales $t_l$ and $t_d$ gives rise to what is known as dissipation length scale $l_d$, defined as $l_d=(\nu/\varepsilon)^{1/4}$ and a similar competition between the time-scales $t_d$ and $t_\O$ allows us to define a length scale $l_{\O d}=\sqrt{(\nu/\O)}$.
Now, lets look at the typical scenario when $Ro$ is moderate.
The four vital length scales are typically arranged according to the order : $l_0>l_\O>l_{\O d}>l_d$.
Thus, the regime $l_0>l>l_\O$ is the regime where effect of rotation is important and anisotropy reigns.
The scales $l\in(l_\O,l_d)$ may be considered to have isotropy, though to be precise, probably $l_d$ here should be replaced by $l_{\O d}$ since rotation seems to be bringing the effect of viscosity to rather larger length scales.
So, now what happens when the $Ro$ is decreased by increasing the angular velocity is interesting.
Both the scales $l_\O$ and $l_{\O d}$ rush towards the dissipation length scale, thereby increasing the anisotropic regime and at the angular velocity $\O=\O_a\equiv\sqrt{(\varepsilon/\nu)}$ one has $l_\O=l_{\O d}=l_d$ and the turbulence is fully anisotropic.
\\
\indent Strictly speaking, even a small rotation introduces anisotropy (however small) at all scales and the isotropic regime does have a degree of anisotropy in it as we shall see in this paper.
In the fully anisotropic limit, {\it i.e.} for $\O=\O_a$, one expects full decoupling of the plane perpendicular to the rotation axis from the direction of the rotation axis.
However, even in the partially anisotropic limit ({\it e.g.} when we have slow rotation imparted on the turbulent fluid), $l_z$ should still be given a special status for being in the direction of the rotation axis, by which we mean that the structure functions should no longer depend on $l$ but rather on $l_z$ and $\vec{l}_\p$ (where $l^2=l_z^2+l_\p^2$ and $|\vec{\O}|=\O_z$).
\\
\indent We shall see how this decoupling sets in, in the limit of low angular velocity and try to study in that very limit, the two-point third order structure function in the first approximation and see how the effect of two-dimensionalisation is all set to sneak in with the switching on of rotation.
\section{$S_3$ for small $\O$}
Let us start with low $\O$-limit.
With this statement we mean, as discussed in the previous section, $\O\ll\O_a$. 
So, the entire fluid may still be treated as isotropic but as rotation should play a role, we assume that $<v_iv_jv'_k>$ (where angular brackets mean ensemble average and $v_i=v_i(\vec{x},t)$ is the $i$-th component of velocity and similarly, $v'_i=v_i(\vec{x}+\vec{l},t)$) should depend on $\vec{\O}$ as well.
$\vec{\O}$ would take care of the mild anisotropy.
Since, physically speaking, $S_3$ should not depend on which way the rotation axis is and since we are interested in low values of $\O$, we shall let $<v_iv_jv'_k>$ depend only on the terms quadratic in $\O$ and not bother about higher order terms in $\O$.
As a result, we write the following most general tensorial form for $<v_iv_jv'_k>$:
\be
b_{ij,k}&\equiv&<v_iv_jv'_k>\nonumber\\
&=&C(l)\delta_{ij}l^o_k+D(l)(\delta_{ik}l^o_j+\delta_{jk}l^o_i)+F(l)l^o_il^o_jl^o_k\nonumber\\
&&+G(l)[(\epsilon_{imk}l^o_j+\epsilon_{jmk}l^o_i)l^o_m]+H(l)\O_i\O_jl^o_k\nonumber\\
&&+I(l)[(\epsilon_{imk}\O_j+\epsilon_{jmk}\O_i)\O_m]\nonumber\\
&&+K(l)(\O_i\O_kl^o_j+\O_j\O_kl^o_i)
\label{2}
\en
where $l_i^o$ is the $i$-th component of the unit vector along $\vec{l}$. We have assumed that the coefficients are dependent only on $l$ and it is the $\vec{\O}$ which is taking care of the mild anisotropy which the turbulent fluid might have.
We must accept that the assumption of letting coefficients depend only on $l$ is rather crude in the light of the complex forms that the two-point tensors in a fully anisotropic turbulence flow take\cite{Biferalee}.
The justification, and hence solace, for the assumption, however, can be drawn from the fact that very simple revealing results matching with recent experiments are arrived at in the long run.
As we are considering incompressible fluid, we must have:
\be
\pa'_kb_{ij,k}=0
\label{3}
\en
which when applied to relation (\ref{2}), yields relationships between various coefficients.
Einstein summation convention has been extensively followed in these calculations unless otherwise specified.
Using relations (\ref{2}) and (\ref{3}), one lands up in the end on the following:
\be
B_{ijk}&\equiv&\la(v'_i-v_i)(v'_j-v_j)(v'_k-v_k)\ra\nonumber\\
&=&2(b_{ij,k}+b_{jk,i}+b_{ki,j})\nonumber\\
&=&-2(lC'+C)(\delta_{ij}l^o_k+\delta_{ik}l^o_j+\delta_{jk}l^o_i)\nonumber\\
&&+6(lC'-C)l^o_il^o_jl^o_k\nonumber\\
&&+4Jl(\O_i\O_jl^o_k+\O_i\O_kl^o_j+\O_j\O_kl^o_i)
\label{4}
\en
Here, prime (``$'$'') denotes derivative w.r.t. $l$ and $J$ is a constant which, curiously enough, is of the same dimension $[L^2T^{-1}]$ as that of the kinematic viscosity.
Now we can see that using the relation (\ref{4}), two-point third order structure function ($S_3$) can be extracted from $B_{ijk}$ in the following way:
\be
S_3(l)&\equiv&\la(\delta v_{\parallel}(\vec{l}))^3\ra\equiv\left\la\left[\left\{\vec{v}(\vec{x}+\vec{l})-\vec{v}(\vec{x})\right\}.\f{\vec{l}}{l}\right]^3\right\ra\nonumber\\
\R S_3(l)&=&\la([v'_i-v_i)l^o_i][(v'_j-v_j)l^o_j][(v'_k-v_k)l^o_k]\ra\nonumber\\
\R S_3(l)&=&B_{ijk}l^o_il^o_jl^o_k\nonumber\\
\R S_3(l)&=&-12C+\f{12J}{l}(\vec{\O}.\vec{l})^2
\label{5}
\en
where we have used relation (\ref{4}).
One may define physical space energy flux ($\varepsilon(\vec{l})$) as:
\be
&&\varepsilon(l)\equiv-\f{1}{4}\vec{\nabla}_l.<|\delta\vec{v}(\vec{l})|^2\delta\vec{v}(\vec{l})>
\label{star}\\
\Rightarrow&&\varepsilon(l)=lC''+7C'+\f{8C}{l}+3J\O^2+\f{6J}{l^2}(\vec{\O}.\vec{l})^2
\label{6}
\en
To get relation (\ref{6}), we have again made use of the relation (\ref{4}).
The energy flux through the wave number $K$ ($\Pi_K$) for the isotropic homogeneous turbulence may be calculated to be:
\be
\Pi_K=\f{2}{\pi}\int_0^\infty dl\f{\sin(Kl)}{l}(1+l\pa_l)\varepsilon(l)
\label{7}
\en
Now if one makes the standard assumption (often made made during the derivation of $S_3$) that as $Re\rightarrow\infty$, the mean energy dissipation per unit mass $\varepsilon(\nu)$ tends to a positive finite value ({\it i.e.,} $\lim_{\nu\rightarrow 0}\varepsilon(\nu)=\varepsilon>0$), then $\lim_{\nu\rightarrow 0}\Pi_K=\varepsilon$ in the inertial regime.
Therefore, in the inertial range, putting $x=Kl$, one has
\be
\Pi_K=\f{2}{\pi}\int_0^\infty dx\f{\sin(x)}{x}f\left(\f{x}{K}\right)=\varepsilon
\label{8}
\en
where,
\be
f\left(\f{x}{K}\right)=f(l)=(1+l\pa_l)\varepsilon(l)
\label{9}
\en
For small $l$ (large $K$), the integral in relation (\ref{8}) yields
\be
f(l)\approx\varepsilon
\label{10}
\en
Now using relations (\ref{6}), (\ref{9}) and (\ref{10}), we form a differential equation which when solved keeping in mind that $S_3$ should not blow up at $l=0$ one gets following form for $S_3$ in slowly rotating homogeneous turbulent fluid. 
\be
S_3(l)=-\f{4}{5}\varepsilon l+\f{12}{5}Jl[\O^2+7(\O_kl_k^o)^2]
\label{11}
\en
One may note from the relation (\ref{11}) that how magically $\O$ has brought up the anisotropic effects even for small $\O$ though for the entire calculation we followed the procedure for the homogeneous isotropic turbulence.
Thus, the form for $S_3$ is pausible.
\\
\indent One may ask: does the effect of two-dimensionalisation shows up in the relation (\ref{11})?
As one may note from the relation (\ref{11}) this is quite a possibility but the only catch being that $J$ should be positive, an issue which we have not been able to resolve.
If $J$ is positive, it means if we increase $\O$ the value of $S_3$ would distort away from the usual $-(4/5)\varepsilon l$ for the non-rotating case to more positive values.
This apparently shows that the effective value of $\varepsilon$ is decreased depicting that the forward energy transfer is depleted which is in keeping with the discussion given in the last paragraph of the section (I) and hence the tendency of the rotating 3D turbulence to show the effect the two-dimensionalisation is being highlighted.
That the sign of $J$ should be positive is a question remains to be addressed.
\\
\indent By the way, the relation (\ref{11}) also suggests that the coefficients in the tensorial form for $b_{ij,k}$ should have dependence on $l_z$ and $l_\p$ separately effecting a mild decoupling of directions.
So taking hint from it, we proceed to rewrite $b_{ij,k}$ for slowly rotating 3D turbulent fluid but now introducing anisotropy directly into the coefficients and not letting $\O$ take care of anisotropy explicitly.
Of course, the coefficients will now depend on $\O$.
\\
\indent For completely isotropic homogeneous turbulence, one would write following general form (relation (\ref{1000})) for $b_{ij,k}$ which is made up of Kronecker delta and components of the unit vectors $\vec{l}/|\vec{l}|$.
\be
b_{ij,k}&=&C(l)\delta_{ij}l^o_k+D(l)(\delta_{ik}l^o_j+\delta_{jk}l^o_i)\nonumber\\
&&+F(l)l^o_il^o_jl^o_k
\label{1000}
\en
The expression is symmetric in $i$ and $j$ and the coefficients are dependent on $l$ only.
As discussed earlier, with rotation coming into effect, anisotropy comes into effect.
If this effects in the possible decoupling (even if partial) of the direction along the rotation axis (which we shall take along the z-axis), then mathematically we may introduce this effect by modifying the form (\ref{1000}) of $b_{ij,k}$ to the following:
\be
b_{ij,k}&=&C(l,l_z,\O)\delta_{ij}l^o_k+D(l,l_z,\O)(\delta_{ik}l^o_j+\delta_{jk}l^o_i)\nonumber\\
&&+F(l,l_z,\O)l^o_il^o_jl^o_k
\label{12}
\en
If one uses the incompressibility condition (relation (\ref{3})), one gets:
\be
&&D=\f{l}{2}(-C'-\f{\dot{C}l_z}{l})-C
\label{50}\\
\textrm{and}\phantom{xxx}&&\dot{D}=0
\label{60}
\en
where dot represents the derivative w.r.t. $l_z$ and prime, as before, the derivative w.r.t. $l$.
Using equation (\ref{50}) in the equation (\ref{60}), one land up on:
\be
&&\ddot{C}l_z+l\dot{C}'+3\dot{C}=0\nonumber\\
&\Rightarrow&C=\sum_n A_nl^{-n-2}l_z^n\nonumber\\
&\Rightarrow&C\ne 0\phantom{x}\textrm{for}\phantom{x}n\in(-\infty,-2]\cap[0,\infty)\nonumber\\
&\Rightarrow&C=D=F=0
\label{70}
\en
In arriving at the result (\ref{70}), we have taken care of the fact that $C$ can not be allowed to blow up for either for $l_z=0$ or for $l=0$.
Thus the relation (\ref{12}) vanishes trivially.
So, we are left with the following choice:
\be
b_{ij,k}&=&C(l_\p,l_z,\O)\delta_{ij}l^o_k+D(l_\p,l_z,\O)(\delta_{ik}l^o_j+\delta_{jk}l^o_i)\nonumber\\
&&+F(l_\p,l_z,\O)l^o_il^o_jl^o_k
\label{13}
\en
Using relation (\ref{3}) and relation (\ref{13}), we arrive at following relationship between the coefficients:
\be
D=-\f{l_\p}{2}\td{C}-\f{l_z}{2}\dot{C}-C\\
F=\f{l^2}{2}\td{\td{C}}+\f{l^2l_z}{2l_\p}\dot{\td{C}}+\left(\f{3l^2}{2l_\p}-\f{l_\p}{2}\right)\td{C}-\f{l_z}{2}\dot{C}-C
\label{14}
\en
Here tilde and dot define derivatives w.r.t. $l_\p$ and $l_z$ respectively.
Proceeding monotonously as before we get
\be
B_{ijk}&=&2(b_{ij,k}+b_{jk,i}+b_{ki,j})\nonumber\\
&=&-2(l_\p\td{C}+l_z\dot{C}+C)(\delta_{ij}l^o_k+\delta_{ik}l^o_j+\delta_{jk}l^o_i)\nonumber\\
&&+6Fl^o_il^o_jl^o_k
\label{15}
\en
And hence,
\be
S_3=B_{ijk}l^o_il^o_jl^o_k=6[F-(l_\p\td{C}+l_z\dot{C}+C)]
\label{16}
\en
The definition for the physical space energy flux ($\varepsilon(\vec{l})$) has to be obviously modified to:
\be
<|\delta\vec{v}(\vec{l})|^2\delta\vec{v}(\vec{l})>
=B_{ii\alpha}l^o_\alpha\f{\vec{l}_\p}{l_\p}+B_{iiz}l^o_z\f{\vec{l}_z}{l_z}
\label{17}
\en
where $\alpha$ takes two values -- $x$ and $y$ only.
Now, using relations (\ref{star}), (\ref{14}), (\ref{15}) and (\ref{17}) and performing tedious algebra one gets:
\be
\varepsilon(l_\p,l_z)&=&\f{-1}{4(l_\p^2+l_z^2)^2}\left[(3l_\p^6+6l_\p^4l_z^2+3l_\p^2l_z^4)\td{\td{\td{C}}}\right.\nonumber\\
&&+\left(3l_\p^5l_z+6l_\p^3l_z^3+3l_\p^4l_z^2+6l_\p^2l_z^4+3l_\p l_z^5\right.\nonumber\\
&&\left.+3l_z^6\right)\dot{\td{\td{C}}}+\left(3l_\p^3l_z^3+6l_\p l_z^5+3l_\p^{-1} l_z^7\right)\ddot{\td{C}}\nonumber\\
&&+\left(5l_\p^5+6l_\p^4 l_z+23l_\p^3 l_z^2+12l_\p^2 l_z^3+18l_\p l_z^4\right.\nonumber\\
&&\left.+6l_z^5\right)\td{{\td{C}}}+\left(-7l_\p^4 l_z+5l_\p^3 l_z^2-l_\p^2 l_z^3\right.\nonumber\\
&&\left.+23l_\p l_z^4+6l_z^5+18l_\p^{-1} l_z^6\right)\dot{\td{C}}+\left(-12l_z^4\right.\nonumber\\
&&-8l_\p^3 l_z-20l_\p^2 l_z^2+36l_\p l_z^3+18l_z^4\nonumber\\
&&\left.+8l_\p^{-1} l_z^5\right)\td{C}+\left(-13l_\p^3 l_z-43l_\p^2 l_z^2-39l_\p l_z^3\right.\nonumber\\
&&\left.\left.-17l_z^4\right)\dot{C}+\left(-4l_\p^3-8l_\p^2l_z-12l_\p l_z^2\right)C\right]\nonumber\\
\label{18}
\en
The energy flux ($\Pi_K$) through the wave number $K$ for the homogeneous (not necessarily isotropic) turbulence may be shown to be:
\be
\Pi_K=\f{1}{2\pi^2}\int_{\m{R}^3}d^3l\f{\sin(Kl)}{l}\vec{\nabla}_l.\left[\varepsilon(\vec{l})\f{\vec{l}}{l^2}\right]
\label{19}
\en
Using cylindrical polar coordinates we reduce the relation (\ref{19}) to:
\be
\Pi_K&=&\f{1}{\pi}\int\int l_\p dl_\p dl_z\left\{\f{\sin(Kl)}{l}\times\right.\nonumber\\
&&\left.\phantom{xxxxxxx}\left[\f{l_\p}{l^2}\f{\pa}{\pa l_\p}+\f{l_z}{l^2}\f{\pa}{\pa l_z}+\f{1}{l^2}\right]\varepsilon(\vec{l})\right\}
\label{20}
\en
Now, we introduce the variables $y=Kl_\p$ and $z=Kl_z$ in relation (\ref{20}) to get:
\be
\Pi_K=\f{1}{\pi}\int_{z=-\infty}^{\infty}\int_{y=0}^\infty dydz\f{\sin(y^2+z^2)^{\f{1}{2}}}{y^2+z^2}\left[f\left(\f{y}{K},\f{z}{K}\right)\right]\nonumber\\
\label{21}
\en
Now, let's probe small $l$ behaviour.
Because $\int_{z=-\infty}^{\infty}\int_{y=0}^\infty dydz[\sin(y^2+z^2)^{1/2}]/(y^2+z^2)=\pi^2/2$, we have
\be
f(l_\p,l_z)\approx \f{2\varepsilon}{\pi}
\label{22}
\en
Obviously, $\varepsilon$ has the meaning of finite positive mean rate of dissipation of energy per unit mass. 
Using the expressions (\ref{18}) and (\ref{22}), we look for the $l_z=0$ limit.
One then has the result:
\be
&&\left[l_\p\f{\pa}{\pa l_\p}+1\right](3l_\p^2\td{\td{\td{C}}}+5l_\p\td{\td{C}}-12\td{C}-4\f{C}{l_\p})=-\f{8\varepsilon}{\pi}\nonumber\\
\Rightarrow&&3l_\p^4\td{\td{\td{\td{C}}}}+14l_\p^3\td{\td{\td{C}}}-2l_\p^2\td{\td{C}}-16{l_\p}\td{C}=-\f{8\varepsilon}{\pi}l_\p\nonumber\\
\Rightarrow&&C=\left(A_1+A_2l_\p^{-1}+A_3l_\p^{\f{7-\sqrt{97}}{6}}+A_4l_\p^{\f{7+\sqrt{97}}{6}}\right)\nonumber\\
&&\phantom{xxxxx}+\f{\varepsilon l_\p}{2\pi}
\label{100}
\en
Relations (\ref{14}), (\ref{16}) and (\ref{100}) together yield following expression for $S_3$:
\be
&&S_3|_{l_z=0}=-\f{6}{\pi}\varepsilon l_\p+A_4\left[3\left(\f{7+\sqrt{97}}{6}\right)\left(\f{1+\sqrt{97}}{6}\right)\right.\nonumber\\
&&\left.\phantom{xxxxxxxxxxxxxx}-12\right]l_\p^{\f{7+\sqrt{97}}{6}}\nonumber\\
\Rightarrow&&S_3|_{l_z=0}=-\f{6}{\pi}\varepsilon l_\p+Al_\p^{\f{7+\sqrt{97}}{6}}
\label{23}
\en
where, $A$ is a constant which for obvious reason depends on $\O$ and $\varepsilon$.
Using dimensional arguments and introducing a non-dimensional constant $c$, we may set
\be
A=c\O^{\f{1+\sqrt{97}}{4}}\varepsilon^{\f{11-\sqrt{97}}{12}}
\label{24}
\en
From relations (\ref{23}) and (\ref{24}), we may write finally
\be
S_3|_{l_z=0}=-\f{6}{\pi}\varepsilon l_\p+c\O^{\f{1+\sqrt{97}}{4}}\varepsilon^{\f{11-\sqrt{97}}{12}}l_\p^{\f{7+\sqrt{97}}{6}}
\label{25}
\en
This (relation (\ref{25})) is the final form for two-point third order structure function in the plane whose normal is parallel to the rotation axis for slowly rotating homogeneous 3D turbulence.
\section{Energy spectrum for small $\O$}
If we for the time being forget about the issue of anomalous scaling, then a mere inspection of the relation (\ref{25}) from the point of view of dimensional analysis would tell that in the directions perpendicular to the axis of rotation, there are two possible energy spectrums {\it viz.}
\be
&&E(k)\sim k^{-\f{5}{3}}
\label{26}\\
\textrm{and}, \phantom{xxx}&&E(k)\sim k^{-\f{16+\sqrt{97}}{9}}
\label{27}
\en
which are respectively due to the first term and the second term in the R.H.S. of the relation (\ref{25}).
It is very interesting to note that the exponent of $k$ in the relation (\ref{27}), {\it i.e.} $-(16+\sqrt{97})/9$, equals $-2.87$ which is in between $-3$ (for 2D turbulence) and $-2$ (for rapidly rotating 3D turbulence as proposed by Zhou).
Obviously, the spectrum (\ref{26}) will be dominant compared to the spectrum (\ref{27}).
But as the $\O$ is increased (of course, remaining within a range so that the anisotropy is not strong enough to breakdown the arguments used to calculate the $S_3$ of the relation (\ref{25})), the spectrum (\ref{27}) becomes more and more prominent; thereby two-dimensionalisation of the 3D homogeneous turbulent fluid is initiated which then carries over to high rotation regime as is being extensively studied.
This signature of two-dimensionalisation is, of course, in agreement with what present literature on turbulence hails as the two-dimensionalisation of turbulence.
\section{Conclusion}
To conclude, we emphasis on the fact that the form of two point third order structure function in a slowly rotating homogeneous 3D turbulence can strongly hint towards the initiation of the effect of two-dimensionalisation of 3D turbulence.
It barely needs to be mentioned that the relations are quite interesting and pertinent (at least within the approximations made in the calculations) -- something which is worth getting in the literature of turbulence since exact relations are very few therein.
So any theory developed in the limit of $Ro\rightarrow 0$ and $Re\rightarrow \infty$, must satisfy the relation derived in this paper in the limit of low $\O$ or explicitly violate the assumptions made to arrive at the result; in this sense the relation (\ref{25}) may prove to be of high importance.
\acknowledgements
CSIR (India) is gratefully acknowledged for awarding fellowship to SC, one of the authors.
Mr. Ayan Paul is thanked for providing the authors with many relevant scholarly articles.


\begin{thebibliography}{99}
%
\bibitem{Batchelor} G. K. Batchelor, {\it An Introduction to Fluid Dynamics}, (Cambridge University Press)
%
\bibitem{Cambon1} C. Cambon and L. Jacquin, { J. Fluid Mech. {\bf {202}}}, 295(1989)
%
\bibitem{Waleffe} F. Waleffe, {Phys. Fluids A {\bf{5}}}, 677(1993).
%
\bibitem{Smith} L.M. Smith, J.R. Chasnov and F. Waleffe, {Phys. Rev. Lett. {\bf 77}}, 2467(1996)
%
\bibitem{Waite} M.L. Waite and P. Bartello, {J. Fluid Mech. {\bf {568}}}, 89(2006)
%
\bibitem{Lindborg} E. Lindborg, {Geophys. Res. Lett. {\bf 32}}, L010809(2005)
%
\bibitem{Baroud1} C. N. Baroud, B. B. Plapp, Z.-S. She and H. L. Swinney, {Phys. Rev. Lett. {\bf 88}}, 114501(2002)
%
\bibitem{Baroud2} C. N. Baroud, B. B. Plapp, H. L. Swinney and Z.-S. She, {Phys. Fluids {\bf 15}}, 2091(2003)
%
\bibitem{Morize1} C. Morize, F. Moisy and M. Rabaud, {Phys. Fluids {\bf 17}}, 095105(2005) 
%
\bibitem{Morize2} C. Morize and F. Moisy, {Phys. Fluids {\bf 18}}, 065107(2006)
%
\bibitem{Zhou} Y. Zhou, {Phys. Fluids {\bf 7}}, 2092(1995)
%
\bibitem{Yeung} P. K. Yeung and Y. Zhou, {Phys. Fluids {\bf 10}}, 2895(1998) 
%
\bibitem{Hattori} Y. Hattori, R. Rubinstein and A. Ishizawa, {Phys. Rev. E {\bf 70}}, 046311(2004)
%
\bibitem{Reshetnyak} M. Reshetnyak and B. Steffen, arXiv:physics/0311001
%
\bibitem{Muller} W.C. Muller and M. Thiele, arXiv:physics/0612207
%
\bibitem{Yang} X. Yang and J. A. Domaradzki, {Phys. Fluids {\bf 16}}, 4088(2004)%
\bibitem{Bellet} F. Bellet, F. S. Godeferd, J. F. Scott and C. Cambon,  ``Wave-turbulence in rapidly rotating flows'',  in {\it Advances in Turbulence X}, edited by H. I. Andersson and P. A. Krogstad, 10th European Turbulence Conference CIMNE, Barcelona, (2004)
%
\bibitem{Galtier} S. Galtier, {Phys. Rev. E {\bf 68}}, 015301(R)(2003)
%
\bibitem{Cambon2} C. Cambon, R. Rubinstein and F. S. Godeferd, {New J. Phys. {\bf 6}}, 73(2004)
%
\bibitem{Frisch} U. Frisch, {\it Turbulence: The Legacy of A. N. Kolmogorov}, (Cambridge University Press)
%
\bibitem{Gawedzki} K. Gawedzki, arXiv:hep-th/9710187
%
\bibitem{Kraichnann} R.H. Kraichnan, {Phys. Fluids {\bf 10}}, 1417(1967)
%
\bibitem{Sagar} S. Chakraborty, arXiv:cond-mat/0611520
%
\bibitem{Bernard} D. Bernard, {Phys. Rev. E {\bf{60}}}, 6184(1999)
%
\bibitem{Biferalee} L. Biferale and I. Procaccia, {Phys. Rep. {\bf 414}}, 43(2005)
\end{thebibliography}
\end{document}